\documentclass{article}
\usepackage[utf8]{inputenc}
\usepackage[
backend=biber,
style=numeric,
maxbibnames=99,
]{biblatex}
\usepackage{graphicx}
\usepackage{comment}
\usepackage{xcolor}

\newcommand{\editor}[1]{}

\title{Dark-matter And Neutrino Computation Explored (DANCE) Community Input to Snowmass}
\author{
Roberts, Amy\footnote{amy.roberts@ucdenver.edu}
\and
Tunnell, Christopher\footnote{tunnell@rice.edu}
\and
von Krosigk, Belina\footnote{belina.krosigk@kit.edu}
\and
Anderson, Tyler
\and
Brodsky, Jason
\and
Buuck, Micah
\and
Cartaro, Tina
\and
Cragin, Melissa
\and
Davies, Gavin S.
\and
Diamond, Miriam
\and
Fan, Alden
\and
Higuera, Aaron
\and
Ippolito, Valerio
\and
Jillings, Chris
\and
Kravitz, Scott
\and
Krezko, Luke
\and
Li, Ivy
\and
Monzani, Maria Elena
\and
Ostrovskiy, Igor
\and
Psihas, Fernanda
\and
Renshaw, Andrew
\and
Riffard, Quentin
\and
Sander, Joel
\and
Sangiorgio, Samuele
\and
Trappitsch, Reto 
\and
Wright, Dennis
}
\date{October, 2019}

\begin{document}
\maketitle

\section{Abstract}

This paper summarizes the needs of the dark matter and neutrino communities as it relates to computation.  The scope includes data acquisition, triggers, data management and processing, data preservation, simulation, machine learning, data analysis, software engineering, career development, and equity and inclusion.  Beyond identifying our community needs, we propose actions that can be taken to strengthen this community and to work together to overcome common challenges.

\section{Introduction}
\editor{EDITOR: Miriam Diamond}

\begin{figure}
    \centering
    \includegraphics[width=1\textwidth]{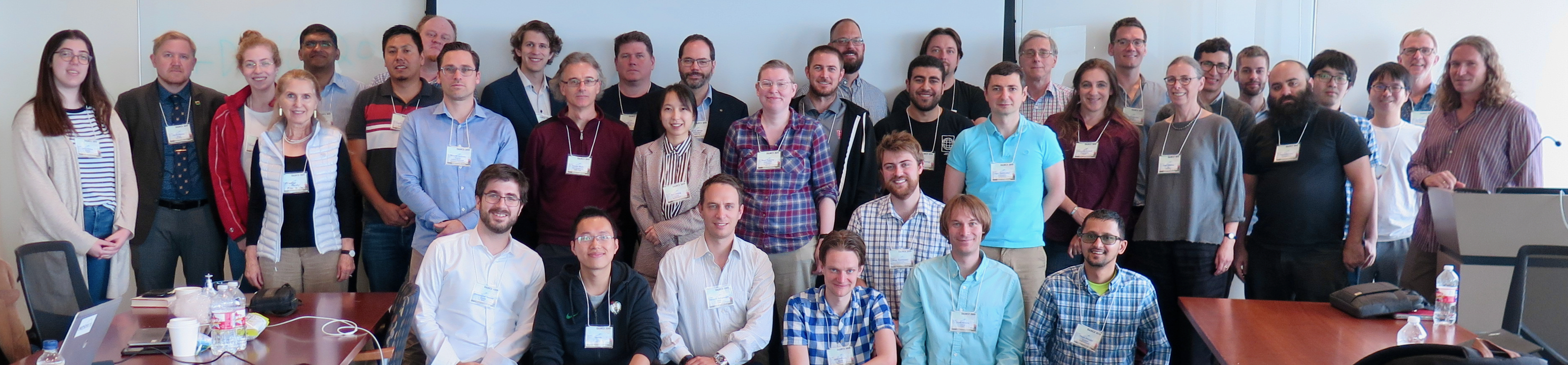}
    \caption{Attendees of the DANCE workshop.}
    \label{fig:dance_photo}
\end{figure}

The Dark-matter And Neutrino Computation Explored (DANCE) community initially came together for a workshop in October of 2019, where this document summarizes the key findings and developments since this time. One of the main goals of the workshop was to identify the challenges related to cyberinfrastructure and researcher training faced by direct-detection dark matter (DM) and medium-scale neutrino experiments. 
Attendees of the workshop represented collaborations spanning multiple countries and totalling thousands of scientists (Ref.~\ref{fig:dance_photo}). 
Particular attention at the workshop was given to identifying common needs and joint efforts to maximize science reach.  
Since this workshop, there have been a range of workshops as part of the SNOWMASS process that have further expanded on these needs~\cite{MariaElenaSnowmass}.

The document is organized as follows. First, a brief description of the science drivers is given. It is followed by sections dedicated to specific computational aspects of the DM and neutrino experiments.  Each section discusses the challenges that experiments share, the communication needs that are specific to the topic, and the current and future states of the field. The document concludes with a summary outcome of the workshop and followup discussions, with overarching recommendations to the community.

\section{Science Drivers}

\textbf{Summary: Dark matter and neutrinos are the most weakly interacting particles, where measuring these requires performing precision rare-event searches for faint signals using a variety of ambitious techniques.}

While there is a wealth of gravitational observations that constrain DM mass density at astrophysical scales, the mass, nature, and mode of interaction of elementary DM particles are still unknown. The DM community responds to these uncertainties by diversifying detector technologies and analyses approaches, which is a feature rather than a flaw.  This includes a wide range of target materials: Si, Ge, Xe, with R\&D proceeding on many others.  There is also a wide range of detection technologies.  All DM experiments collect some version of heat, light, or ionization signal; some of these detectors are sensitive to direction but most are not.

This feature drives a core computing need: smaller collaborations performing R\&D for new types of detectors need support in data acquisition, data management, processing, and analysis library even more than larger-scale experiments.

The low backgrounds, excellent energy resolution, and choice of target materials make many existing DM detectors sensitive double-beta decay instruments. For instance, LZ published a projected sensitivity to the neutrinoless mode of the double-beta decay of $^{136}$Xe~\cite{Akerib:2019dgs}. As another example, XENON1T has measured the half-life of double electron capture of $^{124}$Xe, the longest decay  ever directly observed~\cite{XENON_Collaboration2019-us}.

The current generation of DM experiments have explored new parameter space in axion searches~\cite{Budnik2019-hd,Fu2017-ai,Ahmed2009-jq,XENON_Collaboration2019-us}, fractionally-charged particle searches~\cite{CDMS_Collaboration2015-kg}, and other DM candidates that interact primarily through electron scatter~\cite{Abramoff2019-va, Aguilar-Arevalo:2019wdi, Agnese:HVeV:2018, Agnes:2018oej, Essig:2017aa,Aprile2019-nj}. 

As the next generation of DM detectors turn on, they will produce the lowest-threshold, lowest-radioactivity data sets that have ever been recorded.  As next-generation experiments get closer to the irreducible solar neutrino background, experiments that have directional sensitivity will become increasingly important.  Taking radiopure data at eV-scale energies will allow us to access new information on both neutrinos and dark matter.

The DM community is a collection of many different types of experiments and many detector R\&D efforts that, together, are an unparalleled laboratory for detection of small-signal ionizing radiation.  Delivering science from these facilities requires simulation efforts, data curation, and analysis capabilities. Solving computing problems has always been an isolated effort by each collaboration; as computing challenges increase this isolation places a greater burden on large collaborations and threatens to crush crucial, smaller R\&D efforts.  All-encompassing frameworks have not been successful in this environment, but well-designed, well-tested libraries in any of the computing domains discussed below could serve subsets of the community.  

More generally across science, the science questions that get ask are often restricted when
participation in scientific fields is restricted. To speed up progress towards novel questions and answers we need a more inclusive, less-biased community [citations needed, including Richard Pitt].

\section{Data Acquisition and Triggers}

\textbf{Summary: The sudden increase in experimental data volumes -- up to GB/s -- has required that experiments develop novel and sophisticated techniques for data acquisition and reduction, independently of one another. Common libraries with community-approved APIs would improve DAQ efforts.}


\subsection{Common needs for data acquisition}

Both dark matter and neutrino experiments often have overlaps in their DAQ and triggering requirements, creating opportunities for mutually-beneficial collaboration. 
The explosion of this field (e.g., physics sensitivity) has been matched with an explosion in data rates, numbers of channels, and numbers of users --- making DAQ development increasingly challenging.  
However, individual experiments end up reimplementing solutions as they are often unable to benefit from prior development within the community.  
Having a place to discuss common needs in hardware and data acquisition would make it possible to identify APIs usable by collaborations at large.  APIs could be helpful at both the software layer and the hardware-software interface layer.  The large variance in the types of detector technologies in our field makes it infeasible to create a single framework to support the entire community, but well-publicized libraries with APIs vetted by multiple collaboration could reduce the amount of redundant work collaborations currently pay for. 
Areas that seem promising for such library development are compression, DAQ control, interfacing hardware, and monitoring.  
DAQ collaborations focused on creating libraries would enable novel cyberinfrastructure to permeate throughout the field, as well as any successor technologies.

The major need in this area is to leverage the existing expertise in the DM and neutrino communities and to increase cross-collaboration. 
Experiments need more DAQ support to more efficiently achieve their science objectives, including specifically these take-away points brought up at the workshop:
\begin{itemize}
    \item Experiments need access to well-supported DAQ libraries.
    \item There is a need for scientists that are trained as software developers and are fluent in both analysis and DAQ. This has become more pressing as experiments move to software triggers.
    \item More support is needed for training scientists in hardware techniques, such as FPGA programming.
    \item Additional opportunities for knowledge transfer would be welcome, both between dark matter and neutrino collaborations and outside the field (e.g., nuclear physics, gravitational waves).
    \item Likewise, opportunities are needed for university groups to collaborate with research institutions and industry partners on hardware, firmware, and software.
    \item Usable drivers for off-the-shelf hardware.  Drivers don't always exist and when they do, are not always translatable for the necessary requirements.  Well written drivers for scientific DAQ could help everybody out even if they have different software / data evaluation requirements.
\end{itemize}

\subsection{Data acquisition: communication needs}
Experiments have highly-customized DAQ needs and we do not anticipate sharing full data-acquisition systems.  However, communication about supported DAQ platforms (such as MIDAS) and methods (such as FPGA programming) would be helpful as DAQ expertise is limited.  DAQ training needs are also similar across experiments.  

\subsection{Data acquisition: the current and future landscape}
Data acquisition (DAQ) requirements in the dark matter and neutrino community are driven by our science goals.  
All experiments work to maximize their sensitivity to signal, which typically means building a detector with as much target mass as possible and/or pushing their trigger threshold as low as possible.  

Because we expect the dark-matter cross section to increase as $A^2$ [cite], experiments using lighter isotopes like argon must use more mass than their heavier peers to achieve the same sensitivity.  The increase in mass requires that the DAQ be able to manage the corresponding increase in trigger rate from radioactive isotopes in the detector volume itself and additional interactions with the target mass.

Because we do not know the mass of the dark matter particle, experiments also have the difficult job of designing their detectors for any possible mass.  Many experiments are aiming for sensitivity to dark matter with small masses (sub-GeV) and this directly translates to pushing the trigger threshold as low as possible [cite].  Low thresholds are also important to the neutrino community; detecting solar neutrinos from processes like the p-p interaction rely on thresholds below $\sim 400$\,keV.

Further complicating things, many dark-matter search experiments and neutrino experiments operate in mines to escape the high rates of low-energy neutrons (which can mimic both dark matter and neutrino signal) found near the surface.  This places constraints on data rates.  For example,  SNOLAB's  connection to the surface can handle less than 10 gb/s because it is shared by multiple experiments.  There is no space for many storage racks on-site and the connection to the surface sometimes goes down and experiments must be able to run for at least a day without dumping data.

The current generation of DM and neutrino experiments is moving toward detectors with large masses and hundreds of channels (to thousands in the next generation), making it necessary to deploy complex strategies for online event selection and reconstruction. This presents both technological challenges and opportunities that are new to the community. For example, DEAP's trigger performs hit finding and signal processing (feature extraction) to avoid saving a full digitized waveform. By using such online pulse shape analysis and only storing fit parameters for simple pulses, DEAP is able to nearly halve their data.  
The ultimate example of this would be DarkSide-20k, where one of the key technological hurdles is related to triggering given their choice of target and, therefore, large detector volume.  

The firmware level of the DAQ presents challenges for all dark matter and neutrino experiments. The following trends are currently common in the community:
\begin{itemize}
    \item Most experiments are customizing their existing hardware triggers in collaboration with industry or through detector R\&D.
    \item Most experiments are moving to software triggers for increased scientific flexibility with easier systems.
    \item All experiments are working to understand how these new systems advance analysis and monitoring.
\end{itemize}

Unlike the phase design of the LHC experiments, which are often ASiC-driven, one can now more freely mix hardware, firmware, and software architectures to best achieve the physics goals. Like DEAP, many experiments are working at the firmware level to both improve the physics reach and make their data volumes more manageable. To do so, the experiments often have to collaborate with national labs or industry. Enabling discussions between companies and electronics experts is critical and would help knowledge transfer between collaborations. For example, SuperCDMS has collaborated with Fermilab to implement a noise filter on their hardware trigger in order to lower their threshold as much as possible. 
Similarly, DarkSide is collaborating with CAEN to implement pre-trigger filtering to decrease the rate from SiPM readouts. 

All the DM and neutrino experiments represented at the workshop use software triggers as integral parts of their DAQ. Although these experiments have low backgrounds, most have significant data rates during calibration. For instance, nEXO is anticipating 400 GB/s during calibration runs. DarkSide and PandaX expect comparable data rates. Consequently, most experiments are developing event builder software, some by leveraging existing DAQ frameworks, such as MIDAS (e.g., DEAP, SuperCDMS, and DarkSide), while others develop their own open-source DAQ codes (e.g., XENON1T, XENONnT, PandaX).

As software trigger requirements are strongly tied to physics goals, training software developers to be fluent in both analysis and DAQ is critical. DAQ infrastructure is a necessary part of experiment testing and prototyping. Well-supported DAQ frameworks can significantly ease the burden faced by individual collaborations, as well as drive innovation in detector R\&D for next-generation experiments.

As analysis becomes a part of DAQ through trigger decisions, DM experiments increasingly need to address system design questions that encompass both DAQ and analysis. While DAQ implementations are usually very experiment-specific, most DM experiments have common questions about \begin{it}in situ\end{it} data reduction algorithms, trigger monitoring, and analysis workflows. As such, they would benefit from a way to easily share information. This need for training and for communication between experiments and industry is increasing, as DAQ and trigger systems become more complex for larger, more-sensitive DM experiments. 

\section{Data Management and Processing}

\textbf{Summary: Many experiments in our field have acquired petabytes of data, which is often managed by small teams on the order of an FTE.  This is a new paradigm that is also emerging in other sciences.}

\subsection{Common needs for data management and processing}
Experiments are generating increasing amounts of data -- on the order of peta\-bytes (PBs) per year for the upcoming experiments -- and this requires coordinated data management systems that will scale. There are tools that manage data on this scale, but all of these require effort to adapt these tools to the needs of our community and our specific experiments~\cite{Ahlin:2019iqt}.  Support is needed to adapt these tools to new scientific toolboxes within experiments and related computing clusters (to meet the requirements of the software without adding security risk).  Collaborations must also budget and plan for ongoing support.   

Our experiments require the storage of petabytes of data over the timescale of years, where most of this data will be reprocessed with a frequency ranging from quarterly to yearly, and delay of weeks.  Any data management system needs to support named datasets; some but not all collaborations need to store data across multiple sites.  All collaborations need analysis sites that provide timely data and resource access for all members across an international collaboration.  This can be accomplished with one or more analysis sites depending on the requirements within the collaboration such that users have at most one day of integrated downtime per year.  

Data management tools from HEP exist (e.g., Rucio, DIRAC, GLOBUS) that solve the problem at a larger scale.  Using these tools within the dark matter and neutrino communities runs into three challenges: (1) No central, community-use installation exists.  Experiments must deploy these services at new sites and navigate the security needs of the software and their local clusters.  (2) In contrast to HEP experiments that have used these tools for data storage and analysis for a long time, this is the first time most dark matter and neutrino experiments are using a database-driven data catalog.  Experiments have to determine what metadata needs to be stored and the set of metadata frequently changes as experiments realize the full set of information needed.  (3) HEP tools are built to deliver globally-distributed data to any location in the world with extremely high up time.  Their requirements exceed those of a typical dark matter or neutrino experiment and therefore their software is more complex than is needed by our field.  Collaborations that wish to re-use existing software must also be willing to pay the upfront price of becoming familiar with complex software. 

The key take-away points:
\begin{itemize}
    \item Data management tools from HEP exist (e.g., Rucio, DIRAC, GLOBUS) that solve the problem at a larger scale. The DM and neutrino communities require support to bridge these solutions to their experiments.  
    \item We rely on a single or a small handful of computing sites to provide these services; if these sites change the availability of software services the FTEs required to re-implement starting at a new site are on the order of 2-3 FTEs - something no collaboration has budgeted.  Long-term support plans could help mitigate this, as could efforts such as containerization and grid computing that make computing more site-independent.
    \item Both collaborations and computing clusters need support for ``containers'' and other tools for software distribution to reduce effort spent on adapting processing and management software to run on different clusters. 
    \item Good data management also requires that the community develops a set of best practices around data management plans for small experiments. 
    \item All experiments are struggling with defining the full set of metadata that is needed for reproducible analyses and simulations.  Community discussions about minimum metadata standards would benefit these efforts and increase the ability to re-use existing data resources and infrastructure.
\end{itemize}

\subsection{Data management and processing: communication needs}
Deploying data management and processing tools is site-specific work; we need communication between dark matter experiments to identify software of interest and communication between experiments and HPC administration to make full use of existing knowledge.  Regularized communication will also facilitate the coordination necessary for successful interoperation where software and system processes occur across sites and infrastructure.

Most existing data management and software projects are relatively large and originate outside the dark matter and neutrino fields.  Communication between experiments is critical to clearly defining and advocating for the needs of our community.  Clear pathways to communicate with the developers of these projects is also critical - without clear paths to improve existing software to meet our needs, experiments will continue patching together their own solutions.

\subsection{Data management and processing: the current and future landscape}
As experiments generate data at increasing rates to achieve their science, data management and processing become more important and more complex. The similar challenge of data transport between various production and analysis locations has already been met by LHC experiments. Perhaps, some of what was learned there can be applied in the DM and neutrino community, where an estimated PB of data are handled with on average less than 1 FTE of integrated effort responsible for a range of tasks to adapt existing tools to our field. These tasks include processing data, tracking data, developing the workflow, adapting the workflow with experimental changes, and updating simulations to accept and produce data that is easier to manage.     Adaptations would be required to go the ``last mile'' in
getting data to each group or laboratory in an experiment. These kind of adaptations require personnel who are skilled both in computing and in our science domain - and the need for these people is ongoing as collaborations explore new analyses.  Solving these problems in isolation puts disproportionately large demands on small groups and unnecessary stress on larger groups.     

This section focuses primarily on data tracking, retrieval, and processing.  Data storage is also something collaborations must manage; most partner with a computing site and either buy disks or pay yearly.  SuperCDMS is using the  Open Storage Network (OSN)\cite{Kirkpatrick2021-ob} as a central hub for data access. The OSN is a distributed national service for data storage and transfer at scale, providing ease of access to data that is in active use by projects or communities. 

Examples of tools that may fill data management needs are Rucio, DIRAC, and CKAN. However, these tools were developed by other communities (HEP and the social sciences) and require substantial effort to adapt to our community.  The complexity of these systems places additional burden on every new analyzer in our field.  Both Rucio and DIRAC aim to be comprehensive data management tools for HEP, but it is not clear if the DM and neutrino communities are distributed enough, or large enough, to require them. If only 1 PB of data or less is to be moved, the time lost to deploying, learning, and using a complex tool such as Rucio is a prohibitive burden to collaborations, especially given that dark matter and neutrino collaboration sizes are ten times smaller than typical HEP collaborations. However, there are no widely-used, smaller-scale data management tools that have similar community support.  Collaborations build their own, local solutions ~\cite{Mengel2020-df, datacat:github}.  Any data management system requires ongoing engineering support and the DM and neutrino communities need a solution.  

Another concern about such systems include the degree of conformity required to use them and the possibility that the supporting national labs will change security policies.  Many experiments use GLOBUS to manage data on the cloud, but several national labs are dropping GLOBUS support due to licensing
A long-term support plan from these groups would be useful in our community. In other words, change is likely, so plans must be made to meet it.

Data processing is expected to require high-throughput resources, 
especially if simulation is included. DM and neutrino experiments will increasingly need to process data at multiple facilities to produce timely physics results. Running code at multiple sites poses its own challenges: reproducibility and portability traditionally require time commitment from experienced developers. Tools do exist that can mitigate the extra effort required from collaborations. In particular, support for containers on computing clusters can significantly improve reproducible builds across different systems. Similarly, support for automated workflow software, such as DIRAC and Pegasus, can ensure processing for large amounts of data with limited required time investment from personnel. Successful examples of experiments that use both containerization and workflow software are Project8 and Juno.  Support is required to help DM and neutrino experiments adapt these tools to their use and to enable clusters to support these features.  

Finally, the format of the data itself impacts how much work is required to write and maintain processing code. Data that can be efficiently read in by both HEP-specific and broader tools enables faster prototyping and often provides increased support over experiment-specific tools. Tools such as \texttt{uproot}~\cite{Pivarski2020-nu} and \texttt{Awkward}~\cite{Pivarski2020-qy}, which bridge HEP-specific data formats to standard industry software, are critical. Data-description languages, like DFDL~\cite{DFDL:apache, DFDL:OSG} and Kaitai~\cite{kaitai:website}, could also prove useful at bridging custom formats to standard software.

\section{Data Preservation}

\textbf{Summary: Developing a domain-focused repository of experimental datasets would enable new types of measurements in our field, improve repeatability of existing measurements (including for education), and create standard datasets that engineers can use for designing new algorithms.
Currently, data preservation efforts are isolated  and this greatly increases the risk that useful data will be lost.
}

\subsection{Common needs for data preservation}
Data preservation is necessary for scientific repeatability while  helping maximize the ``return on investment'' of our experiments within science and engineering.  Re-analysis of data is of particular interest, as advances in analysis methods and physics models become available such as more-detailed simulations and new applications of machine learning. Well-understood data sets are often the best source of training for young physicists, and open-access data sets support outreach efforts by targeting both data literacy and science skills. In addition, standardized, trusted data repositories are critical for developing and testing new machine-learning algorithms~\cite{Dua:2019}.  This requires understanding which levels of data and metadata to provide.  The identified needs that are shared across DM and neutrino collaborations are as follows:
\begin{itemize}
    \item Creation of long-term support plans for computing and cyberinfrastructure services for maintaining data on disk and tape, including access management and connection to existing publishing services like Zenodo, which can provide persistent identifiers (e.g. DOIs). 
    \item Development of standards for data release and re-use that are agreed upon by the community.
    \item Investment in metadata and software archiving, which are necessary to support the use of stored data.
\end{itemize}

\subsection{Data preservation: communication needs}
Successful data preservation efforts requires (1) communication across the field, to agree on long-term needs, (2) communication between analysts and computing infrastructure experts, to identify ways to ensure stored data can be analyzed, and (3) communication between the dark matter and neutrino field and stakeholders involved in data-preservation: individual collaborations do not have the resources for long-term data preservation and this is likely a difficult task even for the entire field to shoulder alone.

Finally (4) communication between our field and individuals, agencies, and institution that have a stake in employment and promotion is necessary to ensure that time spent on data preservation is valued.

\subsection{Data preservation: the current and future landscape}
DM and neutrino experiments that wish to preserve data face a number of challenges. Among them is the choice of long-term hosting for different levels of data sets and code, with repository services designed for long-term access and use. In addition, while many collaborations have access to cluster resources, hosting data for next-generation DM experiments will require space on the order of petabytes.  

Unlike library-oriented services such as Zenodo, computing clusters do not typically guarantee hosting for significant lengths of time, nor do they have mechanisms to easily share these data with the public. Additionally, the preserved data must remain usable. Most data in the DM and neutrino communities are stored in custom data formats that require specialized code to analyze. Detailed detector information is often needed, as well as access to metadata such as slow-control and calibration databases. Moreover, community norms for the sharing of data sets are currently absent. Developing community standards can encourage experiments to publish their data and lead to a richer ecosystem of available datasets. In turn, this may lead to an increased use, as people learn to expect data from different experiments to be available to them.

The DM and neutrino communities are not alone in these challenges. More attention to data preservation started in HEP about a decade ago. ICFA (International  Committee  for  Future  Accelerators) formed an inter-collaboration study group that later consolidated into a collaboration. Data from astroparticle experiments supported by NASA are available to the public together with the needed code. The Research Data Alliance~\cite{RDA:website} supports a variety of working groups that specifically address standards for data in different formats to improve reproducibility.  FAIR initiatives span many disciplines in the US and are seeing investment at both the university and national level across the globe~\cite{go-fair:website, Wilkinson2016-ns}.  These efforts often focus on data, but the community is also working to understand the full set of requirements for reproducible science~\cite{Katz2021-pm}.  The dark matter and neutrino community may be able to leverage FAIR efforts for some of the community needs such as data storage and access.  Spaces where existing efforts meet the needs of the community are particularly valuable because of the relatively small size of dark matter and neutrino collaborations.

The DM and neutrino communities can use these fields as guides, but have needs that are not met by existing solutions. For example, open access repositories Zenodo and figshare provide hosting of data sets with sizes up to several gigabytes, while DM experiments may need to store many terabytes of both experimental and simulation data for certain types of studies (e.g. machine learning). The astroparticle community was motivated to establish norms around publishing data sets by a clear mandate from NASA; the DM and neutrino communities likely need similar incentives and resources to make data preservation and access a priority.

\section{Simulation and Physics Generators}

\textbf{Summary: The main backgrounds for direct-detection searches are radiogenic, where predictions rely upon simulation codes without dedicated long-term community support, thereby risking our science program.  Additionally, the software to simulate the complex detectors suffers from similar problems.}

\subsection{Common needs for simulation}
Simulations are an essential component of modern physics experiments. They are vital at every stage of the detector life-cycle, starting with the initial design phase, all the way to the final data analyses. In some cases, the costs to support computing and code development may represent a large fraction of the experiment's budget. Therefore, it is essential to have an efficient, well-maintained, well-understood and thoroughly validated simulation infrastructure. To help achieve this, the following points were identified:
\begin{itemize}
    \item Ensure continuation of Geant4~\cite{ALLISON2016186} support and training within the community.
    \item Increase support and training for detector-specific simulation packages, such as: NEST~\cite{Szydagis_2011,szydagis_m_2018_4262416}, which simulates noble elements detector response, and Opticks~\cite{Simon_blyth_Opticks}, which tracks optical photons.
    \item Provide opportunities for cross-collaboration communication, in order to reduce duplication of effort and maximize return on investment.
    \item Develop computing infrastructure for automated testing and validation of simulation codes.
\end{itemize}

\subsection{Simulation: communication needs}
Communication between simulators and experimenters is essential for useful simulations.  This communication is usually within a collaboration and is usually reasonably effective.

In contrast, communication between experiments with similar simulation needs is currently extremely difficult.  Improving this channel is critical to successful development and uptake of libraries that could benefit the community broadly.

\subsection{Simulation: the current and future landscape}
Dark Matter and neutrino experiments are entering a new era where simulations of both our detectors and background are critical to discovery.  While many simulation needs in the field can be met by existing software packages, challenges include training for parallel and co-processing paradigms and avoiding duplication of effort through a common library of simulation packages across the field.  In addition, physics critical to dark matter and neutrino simulations is sometimes missing from the existing, HEP-centered tools.

Particle transport through materials surrounding low-background detectors is essential to simulate in order to determine the radiation environment for rare searches. Geant4~\cite{ALLISON2016186} has already been mentioned as the most widely used tool for this throughout the community. Improvements in the particle transport are essential for the next generation experiments because they will allow the community to specify the radiation environment for rare searches with the increasing accuracy needed. On the data and computing side of these efforts the community needs to: a) identify physics processes and the phase spaces over which cross section data needs improvement; b) construct a succinct identification system  for \emph{all} physics models present in a given simulation; and c) standardize the data underlying the simulated physics processes.

As all experiments push towards higher sensitivities, understanding backgrounds in terms of detector response becomes increasingly important.  This effort often requires both simulation and experiments, working together. Dark matter and neutrino experiments rely on combinations of light, charge, and heat (phonon) sensors for detection.  This physics often happens at low energies and implementing this physics falls to our community because existing simulation software is designed for higher energies.

In the case of noble liquids, two of the  most computing-intensive challenges are: the simulation of scintillation light with its thousands of optical photons, and the drifting of thermal electrons. The large number of particles to be tracked in these applications is a simulation bottleneck: the passage of one charged particle will generate many low-energy electrons or photons, which in turn also need to be tracked, thus greatly increasing computation time. Offloading the low-energy tracking to co-processors is one way to solve the problem. Another approach is to use parallel or vectorized processing. 

The Opticks package may provide a solution to the tracking of optical photons. Opticks~\cite{Simon_blyth_Opticks,Blyth2021-jb} is a GPU-based ray-tracing code supporting an integration with the Geant4 toolkit. It is estimated to be up to 1,000 times faster than Geant4 alone for the tracking of optical photons. A similar code for thermal electrons does not yet exist, but one could be modeled after Opticks. Reducing the computational cost of a full detector simulation involving optical photons and thermal electrons will lead to a drastic improvement in the level of detail available to our detector model. It will also allow for a better understanding of backgrounds, which are a challenge in rare event searches.

A toolkit used by multiple collaborations is NEST~\cite{Szydagis_2011,szydagis_m_2018_4262416}, which simulates the excitation, ionization, and scintillation processes in noble elements. NEST exists both as a standalone executable and as a callable Geant4 library. Integrating NEST with Opticks would fully leverage the GPU gains in optical photon simulations. A package like NEST is maximally beneficial to the community thanks to its modularity, the quality of its documentation, and the robust plans for long-term support. 

In the case of solid-state detectors, both charge transport and heat (phonons) pose computational challenges because of the large numbers of particles.  G4CMP \cite{brandt2014semiconductor,g4cmp} is a package that provides accurate charge and phonon transport physics to Geant4.  While this package provides the needed physics, further work is necessary to make the code performant so that it can be used to simulate gram-scale detectors.

Full detector response also needs to be simulated, all the way to a DAQ-like format. This task requires generators for light, charge, and heat sensors, cables, and analog and digital electronics. This type of simulation tends to be developed in-house by each experiment (see for example~\cite{Akerib:2020ewf}). But as it is a common need in the field, an increase in code sharing is much desired.  Tools that convert data formats, like DFDL~\cite{DFDL:apache, DFDL:OSG} and Kaitai~\cite{kaitai:website}, could help bridge the gap between standard libraries and experiment-specific data formats.

The Geant4 toolkit underlies most of the simulation frameworks discussed so far.
However, U.S. funding for this project was discontinued in recent years. As a result, the community can no longer rely on core Geant4 developers to address the needs of neutrino and DM experiments. This represents a three-fold challenge: no further updates to the Geant4 engine are expected, including those suggested above; the ENSDF database~\cite{BROWN20181}, which holds the photon evaporation and radioactive decay models of Geant4, will not be updated as often; and finally, user training and support for U.S.-based adopters, which was provided by core Geant4 developers, has all but disappeared.

Commonality of software within the community is highly desirable, as it increases reliability, reduces duplication of effort, and eases the maintenance burden. Several components would aid in this: a shared software repository and continuous integration facility, a core Monte Carlo engine, a collection of physics databases from which the simulation can draw, and a set of event generators that are commonly used by the various experiments. In the current model, each collaboration builds a custom Monte Carlo framework based on Geant4. However, much of this effort is likely to share deep similarities across experiments. The establishment of a common framework would encourage the sharing of new algorithms and physics models. It would also facilitate the validation of the physics output using data from multiple detectors.

\section{Machine Learning} 

\textbf{Summary: Training in relevant libraries and collaboration with machine-learning experts is necessary to maximize the scientific outputs of our experiments.}

\subsection{Common machine learning needs}
As machine learning (ML) grows in use within the DM and neutrino community, work across collaborations is vital to identify and leverage domain-specific applications. Significant effort is required to build confidence in and understanding of its results, so that it can be widely and effectively used. The key needs identified at the workshop toward development of ML in our community are summarized below:
\begin{itemize}
    \item Access to advanced training that is specific to the field, while simultaneously improving opportunities to interact with external experts.
    \item Community-wide efforts to understand uncertainty quantification and physical interpretation of ML results.
    \item Support to make experimental data compatible with external libraries.
    \item Support to produce standard datasets (our own MNIST) to help machine learning practioners understand and co-develop solutions for our problems.
    \item Better access to specialized computing resources, such as GPU and tensor processing unit (TPU) clusters.
\end{itemize}

\subsection{Machine learning: communication needs}
The primary communication needs to effectively use advances in machine learning in the dark matter and neutrino fields are (1) communication between our domain experts and machine-learning experts and (2) training and communication within our field on possible science impact of machine-learning methods.

Because many machine learning libraries expect data in standard formats (like HDF5 or parquet files, or in language-specific data structures like pandas dataframes), communication with data-handling experts is often necessary.

\subsection{Machine learning: the current and future landscape}
Recent advances in ML have enabled new ways to analyze large scientific datasets. The HEP community has successfully leveraged these advances to improve event energy and position reconstruction, particle identification, as well as fast simulation and triggering~\cite{Albertsson2018-dt}. Similar approaches are gaining traction in our community and show promise in addressing some of the challenges facing us moving forward. These include:
\begin{itemize}
    \item Speeding up simulations.
    \item Improving reconstruction and classification by reducing simplifying assumptions inherent in traditional methods.
    \item Optimizing the performance of traditional analysis techniques.
\end{itemize}

Existing DM and neutrino experiments have demonstrated improvements in background rejection using simple neural networks and boosted decision trees~\cite{exo200_excited}. 
Deep learning techniques have also been successfully employed for reconstruction of physical quantities, such as energy and position~\cite{exo_dl,Liang:2021nsz,}, as well as particle identification~\cite{nova:2016,microboone:2017} and signal/background discrimination~\cite{exo_prl}. ML algorithms have also begun to make their way into hardware, through their implementation on FPGAs for triggering~\cite{dune_fpga}. 

Looking beyond these initial successes, using ML to its full potential will require addressing several challenges at the forefront of scientific computing. Core among these is the need to establish ML-based approaches that can match traditional analyses in their reliability and robustness. This is especially relevant for algorithms utilizing low-level information, where building physical intuition about their inner workings is most difficult and the fidelity of simulations is challenging to verify. Techniques to reduce reliance on simulation~\cite{impure_samples} and demonstrate robustness against differences between simulation and data~\cite{minerva} are being explored, but must be further developed and tested.  New techniques leveraging calibration data to generate such data sets would enable expansion of the domain of application of ML in our field.

Quantitative uncertainty estimation in ML outputs is integral to translating results to a final, rigorous statistical analysis. Additionally, tools to improve interpretability of ML outputs can allow human analyzers to benefit from their insights, re-examining their assumptions about what information is physically relevant. More generally, gaining access to state-of-the-art software tools and hardware resources specific to deep learning, such as clusters of graphical and tensor processing units, will help ensure competitiveness with other related fields. Lastly, algorithms currently used in the field are largely based on techniques established in the private technology sector, while our unique problem space may lend itself to approaches that are new to us, such as graphical models. The development of new algorithms, ideally in collaboration with ML experts outside the field of particle physics, has the potential to significantly improve performance on problems that do not readily map onto established paradigms.

To reach these goals, our field must begin training ML experts of our own. General ML resources are readily available online, but are not as efficient as field-specific training. To address this, we recommend advanced dedicated ML workshops to seed some of these activities, where expertise can be shared through interactive tutorials, and promising techniques in the field can be discussed. These should also provide opportunities for collaboration with external experts.  Additionally, we recommend that standard examples are made of the types of data analysis challenges encountered in our field that may lend themselves to machine learning solutions to provide an "MNIST" benchmark for our field.

\section{Data Analysis}

\textbf{Summary: Sophisticated data analysis tools are being iteratively developed to meet the scientific needs of the experiment, without having core libraries such as in other areas of particle physics that they credited with advancing progress on all experiments.}

\subsection{Common data analysis needs}
Dark matter and neutrino experiments all share similar analysis needs.  The core steps of an analysis - which must be performed serially - can be summarized as follows: 

\begin{enumerate}
    \item Use raw sensor data to extract features that are informative about the underlying physical interaction.  Common physical information includes the amount of deposited energy and location of the interaction.
    \item Use the reconstructed data to understand physical processes in the detector, produce calibration constants and develop a detector model.
    \item Remove data that should not be included in a physics analysis, e.g., periods when the detector is too warm, low-frequency noise is high, or cosmic rays have recently passed nearby.
    \item Use the resulting ``clean'' data to model backgrounds and signals in the detector, thereby allowing searches for excess signal events arising from new physical phenomena.
\end{enumerate} 

As the analysis cycle starts with extraction of physics information, there are significant up-front processing needs. Moreover, since dark matter and neutrino detectors are complex, a single analysis must repeatedly iterate these steps. Because these steps must be done in series, when any part in this cycle is slow, be it feature extraction, analysis of physics quantities, or modeling, the time it takes to produce a result may increase dramatically.  This adds significant resource challenges to the analysis procedures, along with the traditional processing and data provenance challenges.  The major issues faced by the dark matter and neutrino communities identified at the workshop are:

\begin{itemize}
    \item Although different dark matter and neutrino experiments all share similar analysis needs, it is difficult to develop analysis software across collaborations because there are almost no common data formats.
    \item Tools that bridge experiment data to high-use, open-source libraries can help the community leverage existing infrastructure and work together on common solutions. Use of such tools has yet to be broadly adopted. 
    \item  As simulation plays an increasingly significant role in understanding detector responses to both signal and background sources, tracking the provenance of information grows in importance, requiring metadata not only for calibration constants and fitting algorithms but also for simulation software. 
    \item The upcoming generation of experiments will see at least an order of magnitude more data, a volume that existing analysis software cannot handle. Tools are critically needed that are capable of interactive analysis of large (of order 100~GB) data sets while staying within a reasonable memory stamp.
\end{itemize}

\subsection{Data analysis: communication needs}
Data analysis is the end goal of experiments, and as such it requires extensive communication between analysts and experts as noted in most of the sections of this paper.  This communication is typically within an experiment but crosses multiple areas of expertise and can therefore be quite challenging, particularly since analyzers are often early-career scientists.

Likewise, analysis expertise is essential across computing activities to ensure that simulations, data acquisition systems, data management, data preservation, and so on are relevant and easily used to support science goals.

Any libraries intended to support multiple experiments must have robust communication between the dark matter and neutrino scientists using the libraries and the developers of any packages.  Ideally, there would be clear contribution and development on-ramps available to the user community.

Finally, communication and training across experiments about analysis methods would benefit the field as there are significant statistical and methodological overlaps between experiments.

\subsection{Data analysis: the current and future landscape}


Many of the challenges mentioned above, e.g, data and software provenance, already have engineering solutions. Approaches like lazy-loading and data warehousing already provide big-data users the ability to work with large sets of data within a reasonable memory stamp. However, these tools and approaches are not ``shovel-ready'' for the DM and neutrino communities.

Provenance tracking may benefit from bridging solutions from the other communities - high-energy physics, geosciences, astronomy, biology and many others have this need - to dark matter and neutrino experiments. As simulation and analysis become increasingly interconnected, fully understanding data provenance requires tracking many types of data simultaneously.  Which database entry was used during a particular analysis or simulation job; who produced that entry, and with what code; which input files, configuration files, and software versions were used?

Interactive analysis is another area where many collaborations share similar challenges and where existing, open-source tools may help solve problems shared by the field. Interactive analysis here refers to exploration and analysis of data on human timescales, in which compute time is smaller than or comparable to the time a human spends deciding the next step. Interactive analysis is critical for the DM community because the detectors are complicated and require significant study before an accurate mapping of signals to physics quantities is possible. The need for interactive analysis has only increased as detectors grow in size, as requirements on backgrounds tighten, and as new analysis techniques like machine learning become available. Supporting interactive analysis becomes more of a challenge as data sets increase in size. The big-data community has a similar problem and has developed tools to lazy-load data that could directly address the issues faced by the DM community. However, these tools are not immediately usable by the community because our data is stored in formats custom to each experiment.

Yet another area that can benefit from existing work is algorithm development, which often requires rapid feedback in a variety of forms. In addition to feedback gained during interactive analysis as described above, visual feedback using graphical event displays is often invaluable in algorithm development. This requires the ability to overlay physics quantities on various stages of the data, including raw sensor data.
Data format also impacts latency of feedback. For instance, NOvA, while working on improving their event reconstruction using deep learning algorithms, changed the data format to the HDF5 standard. This reduced the time frame of their deep learning workflow process from months to days, a 10-fold improvement in the speed of feedback to analyzers.  

In cases like interactive analysis and algorithm development, bridging existing solutions to the DM and neutrino communities will be most valuable if paired with efforts to create small tools that allow collaborations with custom data formats to take advantage of general software. One successful example of such a tool is uproot, which allows collaborations with data stored in ROOT files to work within frameworks such as Numpy and Dask. 

There is a need for a forum for cross-collaboration discussion of shared analysis needs and solutions. The experiences of collaborations with machine learning shows clear value in leveraging existing software infrastructure that enjoys broader support from industry, HEP, or other science disciplines.    

\section{Software Engineering and Developer Tools}
\textbf{Summary: The experimental stacks have increased in complexity, and we don't have FTE, where even using cutting edge technologies hasn't been enough.}

\subsection{Common needs for software development}
The lifetime of DM and neutrino experiments often spans decades, from planning to post-experiment data availability. Such long-term projects generate unique challenges from software engineering, dependency management, and preservation viewpoints. During the planning phase, existing frameworks and tools are usually evaluated for features, performance, and ability to integrate with experiment-specific software. These external dependencies need to be tracked, updated, and deployed throughout the life cycle of the experiment. This adds significant personnel requirements to the project. The tools and approaches to reduce human effort to operate a full software stack are summarized below:

\begin{itemize}
    \item Tools for reproducible software builds, as well as computing cluster support for such tools.
    \item Agreements between collaborations to count work on common tools as work towards collaboration.
    \item Community efforts to test common software packages.
    \item Agreements within collaborations to allow publication of software products. This will allow developers to have peer-reviewed products that are vital to advancement within the field and in industry.
    \item Opportunity to discuss shared needs and identify possible common solutions between collaborations.
\end{itemize}

\subsection{Software development: communication needs}
Communication between experiments is critical to identifying methods that work for our field.  Because personnel for software development is so limited, collaborations often find it difficult to impossible to try new methods; positive reports and implementation expertise from similar collaborations is extremely valuable.

Communication between dark matter and neutrino physicists and HPC resource engineers and managers is also critical; having Singularity~\cite{singularity:website} available across all resources in a collaboration can mean that only one build is needed, whereas any cluster that does not offer this service requires its own plan, maintenance, and maintainer.

Finally, the dark matter and neutrino community needs easy communication with the developers of any packages as we must adapt either libraries or experiment software to fit together.

\subsection{Software development: the current and future landscape}
Software developers within DM and neutrino collaborations spend significant time ``simply'' building software. Several features of long-running experiments make this time-consuming. Firstly, it is common to need to support multiple operating systems during the lifetime of an experiment. For example, the scientific community is currently transitioning from CentOS6 to CentOS7. Most DM and neutrino experiments must build software that works on both systems to make full use of computing resources. Secondly, even within a single experiment, software is written in multiple languages, e.g., C++ and Python. This is often critical to ensure that software is both usable and meets performance requirements, but it further complicates dependency management. Lastly, ensuring the ability to use previous software stacks, or even ensuring that an existing software stack is still usable after an operating system update, requires both a detailed knowledge of dependencies all the way down to individual library versions and control over these dependencies.  Consequently, software within the DM and neutrino communities must be accompanied by information on its dependencies for it to be usable.

The field would benefit from a common tool that helps with building and distributing software, with recipes, documentation, and training resources available on GitHub or other open platforms. Existing build tools do exist: conda~\cite{conda}, Spack~\cite{7832814}, LCGCMake~\cite{Villanueva2019-qs,lcgcmake-gitlab}, EasyBuild~\cite{easybuild}, and Guix~\cite{guix, DBLP:journals/corr/CourtesW15} are all well-supported projects that are designed to ease installation burden and, in some cases, ensure repeatable builds. Another tool, the CernVM File System (CVMFS)~\cite{blomer_jakob_2020_4114078}, aims to easily distribute software. However, transitioning to these systems requires training and upfront time commitments on the part of collaborations and computing clusters. Sparse documentation, a lack of support and training resources, and a lack of cluster support often makes investment in such tools impractical for experiments. Such a system would need to be backed up by continuous integration to verify changes made to recipes, as well as to monitor dependencies between software projects. Furthermore, procedures for deployment to CVMFS, packaging stand-alone containers, or deploying to cloud infrastructure need to be documented in full.

Any such effort would also need recognition by the experiments and funding agencies, such that work on common projects can be accounted as work for the individual experiments. Projects and contributions could be advertised on summary pages, like iris-hep~\cite{Albrecht2019,S2I2HEPSP,irishep} and awesome-hep~\cite{awesome-hep}, made citable via Zenodo.org, and advertised at common DM and neutrino conferences and workshops.

Existing solutions often cover the basic aspects of an experiment’s needs, while custom software needs to be developed to cover the rest. Here it is important to understand both the individual approaches to software engineering and the intersection between similar experiments (e.g. using similar technology or searching for similar physics). The latter can act as a nucleus for more common software, thus lessening the burden on individual experiments. The former would benefit from a coherent training program (see Sec.~\ref{sec:training}), documentation on techniques that worked and techniques that failed, as well as standard recipes for software development pipelines (continuous integration builds, tests, and validation suites).
Existing one-size-fits-all solutions, such as the ART~\cite{Green:2012gv} and GAUDI~\cite{Barrand:2001ny} software frameworks, have great appeal. At the same time, without expertise within an experiment such large, complex frameworks may instead become an obstacle. Therefore, any common software needs to be produced as small, specialized tools to allow for changes and maintainability beyond the lifetime of a single experiment.

\section{Careers, Training, and Personnel}

\label{sec:training}

\textbf{Summary: The DM and neutrino communities do not have enough highly-trained personnel to meet analysis demands.  
We must better understand where gaps in training exist.  Given the increasing importance of scientists who are cross-trained in analysis and software, assessment to identify effective practices is crucial for future experiments.}  

\subsection{Common needs for careers, training, and personnel}
DM and neutrino collaborations need scientists trained in data acquisition, simulation, and analysis at both the user and developer levels to achieve their science goals. Training in these areas has always been critical because of the fast turnover of experts at universities. The need for training in these areas is further heightened in next-generation DM experiments since these areas increasingly overlap: analyzers often need simulation data to complete an analysis, and triggering is often tightly coupled with reconstruction algorithms. This means that scientists must often be proficient in two or more software areas. The following avenues were identified as the most helpful for the dark matter and neutrino fields to train the next generation of scientists:

\begin{itemize}
    \item Creating opportunities for multiple experiments to come together to identify shared training needs and opportunities.
    \item Coordinating with existing training efforts to lessen the burden on the field.
    \item Identifying and supporting opportunities for students to receive in-depth training in collaboration with industry, national labs, and recognized experts.
    \item Developing software that bridges experiment-specific data to common tools, in order to increase training overlap between collaborations.
\end{itemize}

\subsection{Researcher training: communication needs}
Given the significant skill overlap between dark matter and neutrino experiments, the field would greatly benefit from common training efforts.

In addition to cross-experiment communication, open channels between experts outside our field are also critical to successful training efforts: collaboration with experts in educational assessment would help ensure that training efforts are effective and collaboration with other efforts such as FIRST-HEP and Software Carpentry would ensure that we re-use as much existing material as possible and leverage existing awareness.  

Ideally, we would also frequently communicate with industry to align training with hiring practices.

\subsection{Researcher training: current and future landscape}
The DM and neutrino communities share many training needs, including basic programming skills, Geant4 simulation, basic software development skills, familiarity with operating systems and hardware-software interfaces, and rare-event statistical analysis. 
There is currently the DANCE-Edu program run by Rice, U. Denver, and U. Houston, which focuses on developing advanced skills within our field.  Due to the pandemic, this training has either been virtual through YouTube tutorials or relied upon single 'Fellows' that develop a skill that they bring back to a community.  Such activities should be sustained.  Additionally, there are existing training efforts focused on other, more general skills. Software and Data Carpentry workshops train scientists in command-line and programming basics. PyHEP provides training targeted towards physics. Geant4 experts lead workshops aimed at new users. There is some effort in HEP to train scientists in critical skills like DAQ -- the International School of Data Acquisition holds a yearly workshop aimed at data acquisition needs across experiments.  

The DM and neutrino communities do not have enough highly-trained personnel to meet upcoming demands and must work together to identify common training needs and engage with existing training opportunities to better understand where training challenges can be met and what training requires further development.  The above resources are not widely used by the DM and neutrino communities, partly because they do not train students in the methods and frameworks specifically used in our fields. Consequently, training is currently done primarily at the level of individual groups.  This is inefficient, puts unwelcome burden on PIs, and most importantly - creates inequality between groups.  Collaborations struggle to train sufficient personnel for critical software development and maintenance. 

DM and neutrino experiments face a challenge to develop open-source DAQ, simulation, and analysis frameworks and to identify and exploit bridges to existing, broadly-supported tools, such as common data formats (e.g. HDF5) and libraries (e.g. Boost, TensorFlow, Pandas). Using such external tools benefits the career-readiness of students, as their skills are more easily recognized as relevant outside of their immediate field.

In addition, there is a need for partnerships with the curriculum and assessments community in order to identify effective training practices.  Scientists hoping to  develop effective training for the field face a lack of validated assessments for determining the effectiveness of software training at the undergraduate, graduate, and postdoctoral level. While assessments for the K-12 population are increasingly available due to the Next-Generation Science Standards, DM experiments interested in assessing training activities must create their own tools or skip assessment.  This is not a viable solution for upcoming experiments given the importance of scientists who are cross-trained in analysis and software. 

\section{Equity and Inclusion}
\label{sec:equity}

\textbf{Summary: When scientists do not feel welcome, their scientific contributions will suffer, hurting both their careers and the progress of the field at large.  We want to ensure that anybody who is interested in computation within our field does not meet barriers to their success based on their race, gender, or other reasons, therefore we propose centralizing best practices for equity, inclusion, and diversity in collaboration with sociologists who specialize in STEM inclusion and equity.}

\subsection{Common needs to support equity and inclusion in the field}

We include equity and inclusion in this report because (1) the central suggestion of the report is to create an organization to support computing efforts across the dark matter and neutrino community and (2) decisions about the organization scope, funding models, and structure all have an impact on the inclusiveness of our community.  
Building a strong dark matter and neutrino computation community requires diversity, inclusion, and equity work within our sub-field and within physics as a whole.

\begin{itemize}
    \item The scope of the proposed software organization should explicitly include equity and recognize that equity is foundational to the success of our community. Projects focused on equity and inclusion should receive the same funding opportunities as other topics.
    \item Diversifying the dark  matter and neutrino computational community will require efforts to diversify physics in general in addition to work within our specific community~\cite{RichardPitt-convo}.  The scope of the proposed organization is limited to our sub-field, but significantly shifting representation within dark matter and neutrino computing requires collaboration with larger initiatives.
    \item Seek out opportunities to collaborate with experts (such as sociologists) on equity and inclusion in STEM and identify ways to form sustainable partnerships with these experts.  Recognize that these efforts must be funded to be sustainable. 
    \item Ensure that faculty, staff, and trainee positions within our community are viable career options for a diverse group of people.   
\end{itemize}

\subsection{Equity: communication needs}
Communication with experts on Diversity, Equity, and Inclusion (DEI) is critical, given that dark matter and neutrino scientists have limited experience in building programs to improve equity and assessing those programs.  Sustainable partnerships with sociology and existing programs with DEI expertise such as the EDGE mathematics program would greatly increase the chances of success in moving the needle on inclusion in our field.

Communication and cooperation with the broader physics community is also crucial; ``computation in dark matter and neutrino physics`` is too narrow to expect systemic change without effort from our parent community.  

Finally, as with all the other aspects of computing, communication between experiments is crucial.  Computing expertise is already stretched thin and expecting collaborations to create their own, effective efforts from scratch is not practical.

\subsection{Equity: the current and future landscape}
  
Evidence overwhelmingly suggests that under-represented scientists experience isolation and are often explicitly discouraged in physics and computer science at the doctoral, post-doctoral, and faculty levels \cite{black-brown-bruised,Rosa2016-ue,Sharon_Shattuck2020-os,McGee2019-xd,Laursen2020-qy,DAgostino2019-fu}.  Dark Matter and Neutrino computing is a sub-discipline within physics and suffers from the same, dramatic lack of diversity. Addressing these issues increases the science our community can accomplish and therefore belongs in the scope of the proposed organization.

We assert that the scope of this organization must also include equity and inclusion in the broader physics community.  Because the dark matter and neutrino computing community is a sub-field of a sub-field, representation within physics has to improve before we can expect substantial shifts in our area \cite{RichardPitt-convo}.   Including the broader physics community in our scope fits well because computing has become integral to nearly all scientific efforts.  

Creating seismic change requires not only ways to share knowledge and collaborate across experiments, departments, and institutions, but also effective assessment. 

Because scientists working in dark matter and neutrino computation typically have little to no experience with collecting demographic data, there is a significant need to fund partnerships with experts in STEM equity and inclusion.  In addition, partnerships with organizations with this expertise could help: the NSF performs the National Survey of College Graduates, and has just added pilot questions to measure the participation of queer-identified graduates in STEM~\cite{Langin2020-jc}.  The American Institute of Physics is home to the Statistical Research Center, which tracks physics demographics from high school to faculty, as well as employment statistics.  



The DANCE workshop confirmed that a hub organization  for computing and software efforts in the Dark Matter and Neutrino fields is increasingly needed as the community moves towards the next generation of experiments.  Including equity within the scope of this organization is critical to maximizing the science our field can produce.  To maximize impact, our community also needs to find sustainable ways to engage with organizations and experts whose research focus is equity and inclusion in STEM.  Having a meeting place that facilitates this work across experiments and departments enables larger-scale improvements.


\section{Conclusion and Outcomes}
\label{sec:conclusion}

The DANCE workshop was aimed at the community of particle physics experiments focusing on direct-detection dark matter (DM) and medium-scale neutrino physics. This growing community is now approximately one-fifth the size of the collider community, yet so far has made limited investment in cyberinfrastructure to address common needs. The workshop identified common software-related challenges and needs.  Our community provides several overarching recommendations to ensure that future DM and neutrino experimental investments are able to be maximally extracted for scientific results:

\begin{itemize}
    \item Establish new communication pathways that encourage experimental collaborations to leverage prior cyberinfrastructure work, as is done for other technologies our field.
    \item Organize workshops and initiate communication mechanisms to identify technologies and activities, including training, of potential use to the wider community.
    \item Create an entity to stimulate, advocate, and promote across the DM and neutrino communities: synergistic cyberinfrastructure solutions, common software, and training efforts.  
    \item Develop community priorities with one voice such that other entities have a coherent view of our needs.  This is necessary for coherent funding strategies, either budgeted appropriately within existing experiments or forming cross-collaboration projects funded independently.  Additionally,  this will aid interfacing with other communities (LHC, nuclear, astronomy) to maximally leverage their work and share our own novel cyberinfrastructure.
    
\end{itemize}


\textbf{We therefore see a need for a Software Foundation that can serve as a place to incubate efforts across collaborations.}
This could initially entail hosting an annual workshop, such as we did at the DANCE workshop, periodic topical telecons, training, and some common forum for discussion.
There is a clear need for coordination across collaborations in almost every area of computing as every field in science is struggling to adapt to harness the data revolution.

\printbibliography
\end{document}